\documentclass[preprintnumbers,article,amsmath,amssymb,floatfix,10pt,prd,onecolumn,
superscriptaddress,nofootinbib]{revtex4-2}
\usepackage{bm}
\usepackage{amsfonts}
\usepackage{latexsym}
\usepackage[latin1]{inputenc}
\usepackage{graphicx}
\usepackage{amsmath}
\usepackage{palatino}
\usepackage{mathpazo}
\usepackage{textcomp}
\usepackage{float}
\usepackage{booktabs}
\usepackage{dcolumn}
\usepackage{ragged2e}
\usepackage{hyperref}
\hypersetup{colorlinks,citecolor=blue}
\hypersetup{colorlinks=true,linkcolor=red,filecolor=magenta,    urlcolor=blue}
\usepackage{amsmath}
\usepackage{xcolor}
\usepackage{orcidlink}
\usepackage{epsfig}
\usepackage[caption=false]{subfig}
\usepackage{commath}

\usepackage{multirow}
\newcommand{\be}{\begin{equation}}
\newcommand{\ee}{\end{equation}}
\newcommand{\bea}{\begin{eqnarray}}
\newcommand{\eea}{\end{eqnarray}}
\newcommand{\eeas}{\end{eqnarray*}}
\newcommand{\beas}{\begin{eqnarray*}}

\def\jnl@style{\it}
\def\aaref@jnl#1{{\jnl@style#1}}

\def\aaref@jnl#1{{\jnl@style#1}}

\def\aj{\aaref@jnl{AJ}}                   
\def\apj{\aaref@jnl{ApJ}}                 
\def\apjl{\aaref@jnl{ApJ}}                
\def\apjs{\aaref@jnl{ApJS}}               
\def\apss{\aaref@jnl{Ap\&SS}}             
\def\aap{\aaref@jnl{A\&A}}                
\def\aapr{\aaref@jnl{A\&A~Rev.}}          
\def\aaps{\aaref@jnl{A\&AS}}              
\def\mnras{\aaref@jnl{Mon.~Not.~Roy.~Astron.~Soc.}}             
\def\prd{\aaref@jnl{Phys.~Rev.~D}}        
\def\prc{\aaref@jnl{Phys.~Rev.~C}}  
\def\prl{\aaref@jnl{Phys.~Rev.~Lett.}}    
\def\qjras{\aaref@jnl{QJRAS}}             
\def\skytel{\aaref@jnl{S\&T}}             
\def\ssr{\aaref@jnl{Space~Sci.~Rev.}}     
\def\zap{\aaref@jnl{ZAp}}                 
\def\nat{\aaref@jnl{Nature}}              
\def\aplett{\aaref@jnl{Astrophys.~Lett.}} 
\def\apspr{\aaref@jnl{Astrophys.~Space~Phys.~Res.}} 
\def\physrep{\aaref@jnl{Phys.~Rep.}}      
\def\physscr{\aaref@jnl{Phys.~Scr}}       
\def\commat{\aaref@jnl{Comm.~Math.~Phys.}}              
\def\science{\aaref@jnl{Science}}               
\def\cqg{\aaref@jnl{Classical Quant.~Grav.}}            
\def\jpcs{\aaref@jnl{JPCS}}                                     
\def\ijmpd{\aaref@jnl{Int.~J.~Mod.~Phys.~D}}                    
\def\grg{\aaref@jnl{Gen.~Relat.~Gravit.}}               
\def\rpp{\aaref@jnl{Rep.~Prog.~Phys.}}          
\def\npa{\aaref@jnl{Nucl.~Phys.~A}}        
\def\lrr{\aaref@jnl{Living Rev.~Rel.}}                   
\def\jcap{\aaref@jnl{J.~Cosmology Astropart.~Phys.}}    
\def\rmp{\aaref@jnl{Rev.~Mod.~Phys.}}   
\def\epjc{\aaref@jnl{Eur.~Phys.~J.~C}} 
\def\plb{\aaref@jnl{~Phy.~Lett.~B}} 
\def\mpla{\aaref@jnl{Mod.~Phy.~Lett.~A}} 
\def\arxiv{\aaref@jnl{arxiv.org}}

\allowdisplaybreaks[1]

\begin{document}
\color{black}       
\title{Anisotropic background for two fluids: matter and holographic dark
energy}

\author{M. Koussour\orcidlink{0000-0002-4188-0572}}
\email{pr.mouhssine@gmail.com}
\affiliation{Quantum Physics and Magnetism Team, LPMC, Faculty of Science Ben
M'sik,\\
Casablanca Hassan II University,
Morocco.}

\author{M. Bennai\orcidlink{0000-0003-1424-7699}}
\email{mdbennai@yahoo.fr }
\affiliation{Quantum Physics and Magnetism Team, LPMC, Faculty of Science Ben
M'sik,\\
Casablanca Hassan II University,
Morocco.} 
\affiliation{Lab of High Energy Physics, Modeling and Simulations, Faculty of
Science,\\
University Mohammed V-Agdal, Rabat, Morocco.}

\date{\today}

\begin{abstract}
We discuss a spatially homogeneous and anisotropic Bianchi type-I space-time
with two fluids as the content of the Universe: matter and holographic dark
energy in the framework of general relativity. To get the exact solutions of
Einstein's field equations, we choose the scale factor as a hyperbolic
function, specifically, $a\left( t\right) =\sinh ^{\frac{1}{n}}\left( \gamma
t\right) $, where $\gamma $ and $n>0$ are arbitrary constants, which gives
us a time-dependent deceleration parameter. Then we study our cosmological
model under the conditions of the parameters as: $\gamma $ fixed and $n>1$.
Our cosmological solutions led to an early deceleration phase followed by
the current observed acceleration phase. Further, the anisotropic parameter
and some other physical parameters are discussed. We conclude that our
cosmological model is consistent with the results of recent astronomical
observations.
\end{abstract}

\maketitle

\section{Introduction}

\label{sec1}

Recent observations \cite{ref1, ref2, ref3, ref4, ref5, ref6, ref7, ref8,
ref9, ref10, ref11} indicate that our Universe has entered an accelerated
expansion phase. According to Einstein's general relativity (GR), the cause
of such acceleration is the presence of a component of unknown nature,
called dark energy (DE), which has negative pressure and represents $68\%$
of the total density of the Universe, it behaves like a repulsive gravity.
Its nature remains unknown today. It may simply be the cosmological constant 
$\left( \Lambda \right) $ induced by GR which would have a non-zero value.
This cosmological constant has an equation of state (EoS)\ parameter $\omega
=-1$ and is considered to be very consistent with the observation data. In
front of the difficulties linked to its theoretically predicted order of
magnitude with respect to that of the observed vacuum energy \cite{ref12}
other dynamical models of DE have been proposed such as quintessence \cite%
{ref13, ref14}, phantom \cite{ref15}, k-essence \cite{ref16}, tachyons \cite%
{ref17}, Chaplygin gas \cite{ref18}, etc. There another type of DE models,
in which we do not need to introduce any other form of energy, this approach
is called modified gravity theories (MGT), that is, the accelerating
expansion of the Universe can be caused by a modification in gravity.
Moreover, GR is not valid on cosmological scales of matter in the Universe.
The most famous of these theories are: $f(T)$ gravity, $f(G)$ gravity, $%
f(R,G)$ gravity, $f(R,T)$ gravity, $f(R,T,R_{\mu \nu }T^{\mu \nu })$ gravity
and $f(T,T)$ gravity, where $T$ is the trace of the energy-momentum tensor
(or it could be the torsion), $G$ is the Gauss-Bonnet (GB) invariant and $%
R_{\mu \nu }$ is the Ricci tensor \cite{ref19, ref20, ref21, ref22}.

To know the nature of DE, the holographic dark energy (HDE) provides a more
reliable framework for its simplicity and reasonableness. The coincidence
problem can be easily solved for some interactive models of HDE. This model
is considered an application of the holographic principle (HP) to the
problem of DE. The HP was first suggested by G. 't Hooft \cite{ref23} in the
background of black hole physics, then in a cosmological context, another
version of HP was proposed by Fischler and Susskind \cite{ref24}. In the
background of the DE problem, the HP tells us that all physical quantities
in the Universe, including the density of DE $\left( \rho _{\Lambda }\right) 
$, can be described by a few quantities on the boundary of the Universe. It
is clear that it is given in terms of two physical quantities, namely the
reduced Planck mass $\left( M_{p}\right) $\ and the cosmological length
scale $\left( L\right) $ as $\rho _{\Lambda }\approx c^{2}M_{p}^{2}L^{-2}$ 
\cite{ref25}. Next, a relationship was proposed which combines the HDE
density $\left( \rho _{\Lambda }\right) $ and the Hubble parameter $\left(
H\right) $ as $\rho _{\Lambda }\propto H^{2}$, it does not contribute to the
current accelerated expansion of the Universe \cite{ref26}. For purely
dimensional reasons, Granda and Oliveros \cite{ref27} proposed a new
infrared cutoff for the HDE density of the form $\rho _{\Lambda }\approx
\alpha H^{2}+\beta \overset{.}{H}$ where $\alpha $ and $\beta $ are
constants. They show that this new model of DE represents the accelerated
expansion of the Universe and is consistent with current observational data.
Sarkar in several works investigated the HDE in various contexts \cite%
{ref28, ref29, ref30}. In addition, Samanta \cite{ref31} in his work studied
the homogeneous and anisotropic Bianchi type-V Universe filled with matter
and HDE components, and a correspondence between the HDE and quintessence DE
are also established. Recently, Dubey et al. \cite{ref32} Tsallis
holographic dark energy (THDE), infrared cut-off for the Hubble horizon has
been evaluated in the anisotropic Universe using hybrid expansion law (HEL).

The anisotropic Universe has attracted the attention of many researchers
because anisotropy played an important role in the early moments of cosmic
evolution. In addition, the possibility of an anisotropy phase at the
beginning of the Universe followed by an isotropy phase was supported by the
observations. Several researchers have studied homogeneous and anisotropic
Bianchi models, such as the spatially homogeneous and anisotropic Bianchi
type-I model, which is a direct generalization of the FLRW Universe with a
scale factor in each spatial direction \cite{ref33, ref34, ref35}. In this
study, we analyze a spatially homogeneous and anisotropic Bianchi type-I
space-time with two fluids as the content of the Universe: matter and
holographic dark energy in the framework of general relativity. Moreover, to
find the exact solutions of the field equations and some physical
parameters, we assume the scale factor as a hyperbolic function,
specifically, $a\left( t\right) =\sinh ^{\frac{1}{n}}\left( \gamma t\right) $%
, where $\gamma $\ and $n$ are free model parameters, which gives us a
time-dependent deceleration parameter (DP).

The present paper is organized as follows: In Sec. \ref{sec2} we present the
field equations for the Bianchi type-I Universe and defined some physical
and geometrical parameters to solve the field equations in the same section.
In Sec. \ref{sec3}, we solve the field equations by assuming a hyperbolic
function of the scale factor. Finally, in Secs. \ref{sec3} and the last, we
discuss the jerk parameter and conclude our results, respectively.

\section{Metric and basic field equations}

\label{sec2}

In our analysis, we consider a spatially homogeneous and anisotropic Bianchi
type-I metric \cite{ref28}

\begin{equation}
ds^{2}=dt^{2}-A^{2}\left( t\right) dx^{2}-B^{2}\left( t\right)
dy^{2}-C^{2}\left( t\right) dz^{2},  \label{eqn1}
\end{equation}%
where $A\left( t\right) $, $B\left( t\right) $, and $C\left( t\right) $ are
the directional scale factors, functions of cosmic time $t$ only. We will
follow the same steps in the literature, and first, write the expressions
for the physical and geometrical parameters that we will use here to solve
Einstein's field equations for the metric of Eq. (\ref{eqn1}).

The average scale factor $a$ of the Bianchi type-I space-time is given by

\begin{equation}
a=\left( ABC\right) ^{\frac{1}{3}}.  \label{eqn2}
\end{equation}

The spatial volume $V$\ of the Universe is defined as%
\begin{equation}
V=a^{3}=ABC.  \label{eqn3}
\end{equation}

Further, the directional Hubble parameters $H_{1}$, $H_{2}$, and $H_{3}$ are
respectively

\begin{equation}
H_{1}=\frac{\overset{.}{A}}{A},\text{ \ \ \ \ }H_{2}=\frac{\overset{.}{B}}{B}%
,\text{ \ \ \ \ }H_{3}=\frac{\overset{.}{C}}{C}.  \label{eqn4}
\end{equation}

The average Hubble parameter is defined as

\begin{equation}
H=\frac{1}{3}\left( H_{1}+H_{2}+H_{3}\right) .  \label{eqn5}
\end{equation}

From Eqs. (\ref{eqn2})-(\ref{eqn5}), we find

\begin{equation}
H=\frac{1}{3}\frac{\overset{.}{V}}{V}=\frac{1}{3}\left( \frac{\overset{.}{A}%
}{A}+\frac{\overset{.}{B}}{B}+\frac{\overset{.}{C}}{C}\right) .  \label{eqn6}
\end{equation}

Other physical parameters, the expansion scalar $\left( \theta \right) $,
average anisotropic parameter $\left( A_{m}\right) $ and shear scalar $%
\left( \sigma ^{2}\right) $, are defined for the Bianchi type-I metric (\ref%
{eqn1}), as

\begin{equation}
\theta =\frac{\overset{.}{A}}{A}+\frac{\overset{.}{B}}{B}+\frac{\overset{.}{C%
}}{C},  \label{eqn7}
\end{equation}

\begin{equation}
A_{m}=\frac{1}{3}\overset{3}{\underset{i=1}{\sum }}\left( \frac{\Delta H_{i}%
}{H}\right) ^{2},  \label{eqn8}
\end{equation}

\begin{equation}
\sigma ^{2}=\frac{1}{2}\left[ \left( \frac{\overset{.}{A}}{A}\right)
^{2}+\left( \frac{\overset{.}{B}}{B}\right) ^{2}+\left( \frac{\overset{.}{C}%
}{C}\right) ^{2}\right] -\frac{\theta ^{2}}{6},  \label{eqn9}
\end{equation}%
where $\Delta H_{i}=H_{i}-H$ and $H_{i}\left( i=1,2,3\right) $ represent the
directional Hubble parameters.

The Einstein's field equation (with $8\pi G=1$ and $c=1$) is given by

\begin{equation}
R_{\mu \nu }-\frac{1}{2}g_{\mu \nu }R=-\left( T_{\mu \nu }+\overline{T}_{\mu
\nu }\right) ,  \label{eqn10}
\end{equation}%
where $R_{\mu \nu }$ is the Ricci tensor, $R$ is the Ricci scalar and $%
T_{\mu \nu }$, $\overline{T}_{\mu \nu }$\ are the energy-momentum tensors of
matter and HDE respectively. These energy-momentum tensors are defined as

\begin{equation}
T_{\mu \nu }=\rho _{m}u_{\mu }u_{\nu },  \label{eqn11}
\end{equation}%
and

\begin{equation}
\overline{T}_{\mu \nu }=\left( \rho _{\Lambda }+p_{\Lambda }\right) u_{\mu
}u_{\nu }+g_{\mu \nu }p_{\Lambda },  \label{eqn12}
\end{equation}%
where $\rho _{m}$, $\rho _{\Lambda }$ are the energy densities of the
matter, respectively, while $p_{\Lambda }$ is the pressure of the HDE.

The Einstein's field equations (\ref{eqn10}), with (\ref{eqn11}) and (\ref%
{eqn12}) for the metric (\ref{eqn1}) leads to the following system of field
equations

\begin{equation}
\frac{\overset{.}{A}}{A}\frac{\overset{.}{B}}{B}+\frac{\overset{.}{B}}{B}%
\frac{\overset{.}{C}}{C}+\frac{\overset{.}{C}}{C}\frac{\overset{.}{A}}{A}%
=\rho _{m}+\rho _{\Lambda },  \label{eqn13}
\end{equation}

\begin{equation}
\frac{\overset{..}{A}}{A}+\frac{\overset{..}{B}}{B}+\frac{\overset{.}{A}%
\overset{.}{B}}{AB}=-p_{\Lambda },  \label{eqn14}
\end{equation}

\begin{equation}
\frac{\overset{..}{B}}{B}+\frac{\overset{..}{C}}{C}+\frac{\overset{.}{B}%
\overset{.}{C}}{BC}=-p_{\Lambda },  \label{eqn15}
\end{equation}

\begin{equation}
\frac{\overset{..}{C}}{C}+\frac{\overset{..}{A}}{A}+\frac{\overset{.}{C}%
\overset{.}{A}}{CA}=-p_{\Lambda },  \label{eqn16}
\end{equation}%
where $\left( \overset{.}{}\right) $\ dot represents a derivative with
respect to cosmic time.

Now, subtracting Eq. (\ref{eqn16}) from Eq. (\ref{eqn15}) we get

\begin{equation}
\frac{d}{dt}\left( \frac{\overset{.}{A}}{A}-\frac{\overset{.}{B}}{B}\right)
+\left( \frac{\overset{.}{A}}{A}-\frac{\overset{.}{B}}{B}\right) \left( 
\frac{\overset{.}{A}}{A}+\frac{\overset{.}{B}}{B}+\frac{\overset{.}{C}}{C}%
\right) =0.  \label{eqn17}
\end{equation}

Using Eq. (\ref{eqn3}) we can write Eq. (\ref{eqn17}) in the form

\begin{equation}
\frac{d}{dt}\left( \frac{\overset{.}{A}}{A}-\frac{\overset{.}{B}}{B}\right)
+\left( \frac{\overset{.}{A}}{A}-\frac{\overset{.}{B}}{B}\right) \frac{%
\overset{.}{V}}{V}=0.  \label{eqn18}
\end{equation}

By integrating the above equation, we get

\begin{equation}
\frac{\overset{.}{A}}{A}=d_{1}\exp \left( k_{1}\int \frac{dt}{V}\right) ,
\label{eqn19}
\end{equation}%
where $d_{1}$ and $k_{1}$\ are constants of integration.

Similarly, subtracting Eq. (\ref{eqn14}) from Eq. (\ref{eqn13}) and Eq. (\ref%
{eqn13}) from Eq. (\ref{eqn15}) we find

\begin{equation}
\frac{\overset{.}{B}}{B}=d_{2}\exp \left( k_{2}\int \frac{dt}{V}\right) ,
\label{eqn20}
\end{equation}

\begin{equation}
\frac{\overset{.}{C}}{C}=d_{3}\exp \left( k_{3}\int \frac{dt}{V}\right) ,
\label{eqn21}
\end{equation}%
where $d_{2}$, $d_{3}$, $k_{2}$ and $k_{3}$\ are constants of integration.
From Eq. (\ref{eqn3}) one can obtain the the relation between the constants $%
d_{1}$, $d_{2}$, $d_{3}$, $k_{1}$, $k_{2}$, and $k_{3}$\ as $%
d_{2}=d_{1}d_{3} $, $k_{2}=k_{1}+k_{3}$.

From Eqs. (\ref{eqn19})-(\ref{eqn21}), the directional scale factors $%
A\left( t\right) $, $B\left( t\right) $ and $C\left( t\right) $ can be
explicitly written in terms of the the average scale factor $a\left(
t\right) $ as

\begin{equation}
A(t)=l_{1}a\exp \left( m_{1}\int a^{-3}dt\right) ,  \label{eqn22}
\end{equation}

\begin{equation}
B(t)=l_{2}a\exp \left( m_{2}\int a^{-3}dt\right) ,  \label{eqn23}
\end{equation}

\begin{equation}
C(t)=l_{3}a\exp \left( m_{3}\int a^{-3}dt\right) ,  \label{eqn24}
\end{equation}%
where $l_{1}$, $l_{2}$, $l_{3}$, and $m_{1}$, $m_{2}$, $m_{3}$ are constants
satisfy the following two relations

\begin{equation}
l_{1}l_{2}l_{3}=1,\text{ \ \ \ \ }m_{1}+m_{2}+m_{3}=0.  \label{eqn25}
\end{equation}

Now by using Eqs. (\ref{eqn13})-(\ref{eqn16}) and the barotropic EoS $%
p_{\Lambda }=\omega _{\Lambda }\rho _{\Lambda }$, we obtained the continuity
equation as

\begin{equation}
\overset{.}{\rho }_{m}+\left( \frac{\overset{.}{A}}{A}+\frac{\overset{.}{B}}{%
B}+\frac{\overset{.}{C}}{C}\right) \rho _{m}+\overset{.}{\rho }_{\Lambda
}+\left( \frac{\overset{.}{A}}{A}+\frac{\overset{.}{B}}{B}+\frac{\overset{.}{%
C}}{C}\right) \left( 1+\omega _{\Lambda }\right) \rho _{\Lambda }=0
\label{eqn26}
\end{equation}

For two fluids anisotropic: matter and HDE, the continuity equation (\ref%
{eqn26}) can be written as

\begin{equation}
\overset{.}{\rho }_{m}+\left( \frac{\overset{.}{A}}{A}+\frac{\overset{.}{B}}{%
B}+\frac{\overset{.}{C}}{C}\right) \rho _{m}=0,  \label{eqn27}
\end{equation}%
and

\begin{equation}
\overset{.}{\rho }_{\Lambda }+\left( \frac{\overset{.}{A}}{A}+\frac{\overset{%
.}{B}}{B}+\frac{\overset{.}{C}}{C}\right) \left( 1+\omega _{\Lambda }\right)
\rho _{\Lambda }=0,  \label{eqn28}
\end{equation}%
respectively.

\section{Cosmological solutions of the model}

\label{sec3}

Taking into account the new form proposed in \cite{ref27}, we assume the
following HDE density for our analysis

\begin{equation}
\rho _{\Lambda }=3\left( \alpha H^{2}+\beta \overset{.}{H}\right) ,
\label{eqn29}
\end{equation}%
where $\alpha $, $\beta $ are constants that must satisfy the constraints
imposed by the present observational data, $H$ is the average Hubble
parameter and $M_{p}^{-2}=8\pi G=1$. This new HDE model for energy density
introduced by \cite{ref27} may be important in comprehension the evolution
of the Universe especially, an anisotropic Universe. As mentioned in the
introduction, the advantage of this model is that it predicts the
accelerated expansion of the Universe and is coherent with the present
observational data. In addition, for the characteristic length scale $L$
(Infrared cut-off) found in the expression of HDE, there are several
possible options in the literature\ such as the Hubble horizon, future event
horizon or particle horizon \cite{Xu}. Recently, Chen and Jing \cite{Chen}
have modified this new HDE model in which energy density of HDE contains the
second order derivative of Hubble's parameter with regard to cosmic time and
named it as modified holographic Ricci dark energy.

Using Eq. (\ref{eqn29}) in Eq. (\ref{eqn28}), the EoS parameter for HDE is
obtained as

\begin{equation}
\omega _{\Lambda }=-1-\frac{2\alpha H\overset{.}{H}+\beta \overset{..}{H}}{%
3H\left( \alpha H^{2}+\beta \overset{.}{H}\right) }.  \label{eqn30}
\end{equation}

To solve these field equations we assume the cosmological scale factor as a
hyperbolic function,

\begin{equation}
a\left( t\right) =\sinh ^{\frac{1}{n}}\left( \gamma t\right) ,  \label{eqn31}
\end{equation}%
where $\gamma $ and $n>0$ are arbitrary constants. The motivation to choose
the scale factor obtained in Eq. (\ref{eqn31}) is that it produces a
time-dependent deceleration parameter. It belongs to a class of models that
describe the transition of the Universe from early decelerated phase to the
recent accelerating phase as indicated by the recent observations in
cosmology. The derivation and the motivation to choose such scale factor has
already been described in details by Chawla et al. \cite{ref36}. The role of
two fluid minimally coupled in the evolution of the dark energy parameter
has been investigated by Pradhan \cite{ref37} with the help of the
hyperbolic solution of the scale factor. Esmaeili and Mishra \cite{Esmaeili}
constructed the cosmological model in $f(R,T)$ theory of gravity in a
Bianchi type VIh Universe by using the hyperbolic scale factor. Recently,
Pradhan et al. \cite{ref38} proposed the hyperbolic form to examine the
physical behaviour of the transition of the anisotropic Bianchi type-I
perfect fluid cosmological models from early decelerating to the current
accelerating phase in the framework of $f\left( R,T\right) $ gravity. In
addition, Singh and Lalke \cite{Singh} studied the background of flat FLRW
metric in the framework of $f(Q,T)$ gravity theory and considered two
cosmological models by taken the parameterization of the scale factor as a
hyperbolic function.

Using Eq. (\ref{eqn31}) for the average scale factor in Eqs. (\ref{eqn22})-(%
\ref{eqn24}), we obtain the directional scale factors of the following form 
\cite{ref39}

\begin{equation}
A\left( t\right) =l_{1}\sinh ^{\frac{1}{n}}\left( \gamma t\right) \exp \left[
\frac{m_{1}\left( -1\right) ^{\frac{n+3}{2n}}}{2\gamma }\cosh \left( \gamma
t\right) F\left( t\right) \right] ,  \label{eqn32}
\end{equation}

\begin{equation}
B\left( t\right) =l_{2}\sinh ^{\frac{1}{n}}\left( \gamma t\right) \exp \left[
\frac{m_{2}\left( -1\right) ^{\frac{n+3}{2n}}}{2\gamma }\cosh \left( \gamma
t\right) F\left( t\right) \right] ,  \label{eqn33}
\end{equation}

\begin{equation}
C\left( t\right) =l_{3}\sinh ^{\frac{1}{n}}\left( \gamma t\right) \exp \left[
\frac{m_{3}\left( -1\right) ^{\frac{n+3}{2n}}}{2\gamma }\cosh \left( \gamma
t\right) F\left( t\right) \right] ,  \label{eqn34}
\end{equation}%
where

\begin{equation}
F\left( t\right) =1+\frac{1}{6}\left( 1+\frac{3}{n}\right) \cosh ^{2}\left(
\gamma t\right) +\frac{3}{40}\left( 1+\frac{3}{n}\right) \left( 1+\frac{1}{n}%
\right) \cosh ^{4}\left( \gamma t\right) +\circ \left[ \cosh \left( \gamma
t\right) \right] ^{6}.  \label{eqn35}
\end{equation}

The directional Hubble parameters $H_{i}$ and the average Hubble parameter $%
H $ become

\begin{equation}
H_{1}=\frac{\gamma }{n}\coth \left( \gamma t\right) +\frac{m_{1}}{\sinh ^{%
\frac{3}{n}}\left( \gamma t\right) },  \label{eqn36}
\end{equation}

\begin{equation}
H_{2}=\frac{\gamma }{n}\coth \left( \gamma t\right) +\frac{m_{2}}{\sinh ^{%
\frac{3}{n}}\left( \gamma t\right) },  \label{eqn37}
\end{equation}

\begin{equation}
H_{3}=\frac{\gamma }{n}\coth \left( \gamma t\right) +\frac{m_{3}}{\sinh ^{%
\frac{3}{n}}\left( \gamma t\right) },  \label{eqn38}
\end{equation}

\begin{equation}
H=\frac{\gamma }{n}\coth \left( \gamma t\right) .  \label{eqn39}
\end{equation}

The expansion scalar $\theta $ and shear scalar $\sigma ^{2}$ are obtained as

\begin{equation}
\theta =3H=\frac{3\gamma }{n}\coth \left( \gamma t\right) ,  \label{eqn40}
\end{equation}

\begin{equation}
\sigma ^{2}=\left( m_{1}^{2}+m_{2}^{2}+m_{3}^{2}\right) \left( \frac{1}{%
\sinh ^{\frac{3}{n}}\left( \gamma t\right) }\right) ^{2}.  \label{eqn41}
\end{equation}

From Eqs. (\ref{eqn39})-(\ref{eqn41}), we can see that the Hubble parameter,
scalar expansion, and scalar shear are diverge at $t=0$ and approach to zero
at $t\rightarrow \infty $. Now, using Eq. (\ref{eqn31}) into Eq. (\ref{eqn3}%
) we get the spatial volume of the Universe as

\begin{equation}
V=\sinh ^{\frac{3}{n}}\left( \gamma t\right) .  \label{eqn42}
\end{equation}

From the above equation, it is clear that the spatial volume of our model
increases exponentially with cosmic time and zero at initial time $t=0$. In
addition, it shows that the evolution of our Universe commences from a big
bang scenario. The average scale factor in Eq. (\ref{eqn31}) is also zero at
the early epoch of the Universe. Therefore, our model has a singularity of
type point \cite{ref40}.

The average anisotropy parameter $A_{m}$ is given as

\begin{equation}
A_{m}=\left( \frac{m_{1}^{2}+m_{2}^{2}+m_{3}^{2}}{3}\right) \left( \frac{n}{%
\gamma \coth \left( \gamma t\right) \sinh ^{\frac{3}{n}}\left( \gamma
t\right) }\right) ^{2}.  \label{eqn43}
\end{equation}

\begin{figure*}[ht]
\centerline{\includegraphics[scale=0.8]{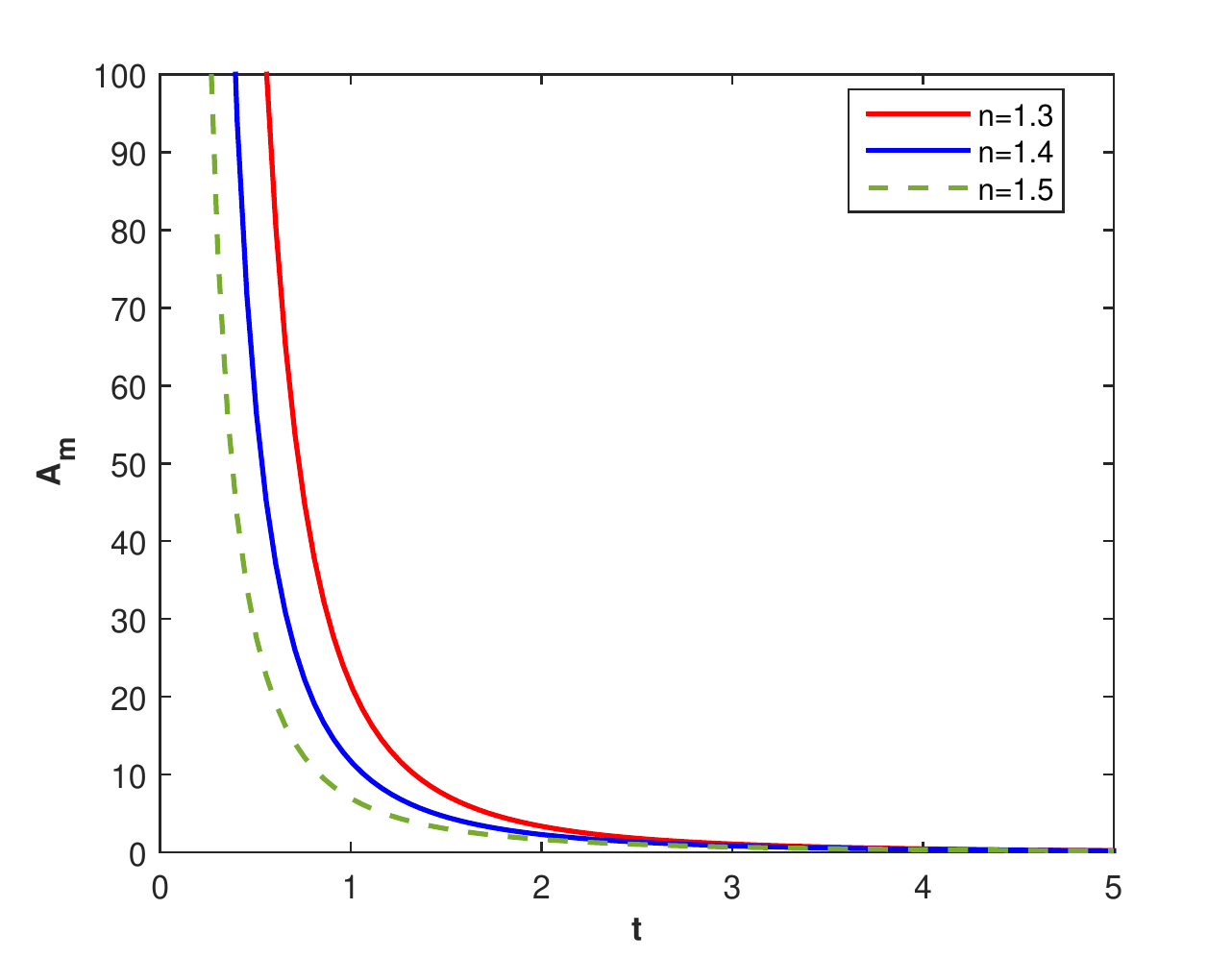}}
\caption{{The plot of the anisotropy parameter $A_{m}$\ vs. time $t$ with $%
\protect\gamma =0.1$, $m_{1}^{2}+m_{2}^{2}+m_{3}^{2}=9.4\times 10^{-4}$.}}
\label{fig1}
\end{figure*}

From Fig. \ref{fig1} it is clear that the average anisotropic parameter $%
A_{m}$ is a decreasing function of cosmic time, which tends towards zero at $%
t\rightarrow \infty $. This indicates that our cosmological model contains a
transition from the early anisotropic Universe to the current isotropic
Universe as DE starts to dominate the energy density of the Universe, this
characteristic is consistent with recent observations. In addition, all
model parameters are chosen based on the constraints imposed by the current
observational data. In the literature, the model parameters are constrained
by using one of the available datasets such as $31$ points of the Hubble
datasets, 6 points of the BAO (Baryon Acoustic Oscillations) datasets and $%
580$ points from type Ia supernovae (SNe Ia). According to the analysis in 
\cite{ref42}, the free parameter $n$ is fit with the observational data. The
constrained value of $n$ are obtained as $1.5176$, $1.5907$, $1.5009$, $%
1.5396$ and $1.5060$ corresponding to the Hubble $H(z)$, SNe Ia, BAO, $H(z)$
+ SNe Ia and $H(z)$ + SNe Ia + BAO datasets.

Several recent observational data have shown that a positive value\ of the
DP $\left( q>0\right) $ describes a decelerating Universe, and a negative
value $\left( q<0\right) $ describes the acceleration of the cosmic
expansion, other observational data from SNe Ia has shown that the current
Universe in the acceleration phase and the value of the DP is confined to
range $-1\leq q<0$. The DP is defined as

\begin{equation}
q=-1+\frac{d}{dt}\left( \frac{1}{H}\right) .  \label{eqn44}
\end{equation}

Using Eq. (\ref{eqn39}), the DP for our cosmological model is

\begin{equation}
q=n\left[ 1-\tanh ^{2}\left( \gamma t\right) \right] -1.  \label{eqn45}
\end{equation}

From Eq. (\ref{eqn45}), we can find the relation between the parameters of
the model $n$ and $\gamma $ as follows

\begin{equation}
\gamma t_{0}=\tanh ^{-1}\left( \frac{n-q_{0}-1}{n}\right) ^{\frac{1}{2}},
\label{eqn46}
\end{equation}%
where $t_{0}$ is the present time and $q_{0}$ is the present value of DP, to
analysis the behavior of certain parameters we consider $t_{0}=13.8GYs$ and $%
q_{0}=-0.54$ \cite{ref41}. In addition, using the relation which connects
the average scale factor and the redshift $a=a_{0}\left( 1+z\right) ^{-1}$,
where $a_{0}$ is the present value of the scale factor, i.e. at $z=0$, we
obtained the following expression

\begin{equation}
t\left( z\right) =\frac{\sinh ^{-1}\sqrt{\frac{n-\left( q_{0}+1\right) }{%
\left( z+1\right) ^{2n}\left( q_{0}+1\right) }}}{\gamma },  \label{eqn47}
\end{equation}

\begin{equation}
H\left( z\right) =\frac{\gamma \coth \left( \sinh ^{-1}\sqrt{\frac{n-\left(
q_{0}+1\right) }{\left( z+1\right) ^{2n}\left( q_{0}+1\right) }}\right) }{n},
\label{eqn48}
\end{equation}

\begin{figure*}[ht]
\centerline{\includegraphics[scale=0.8]{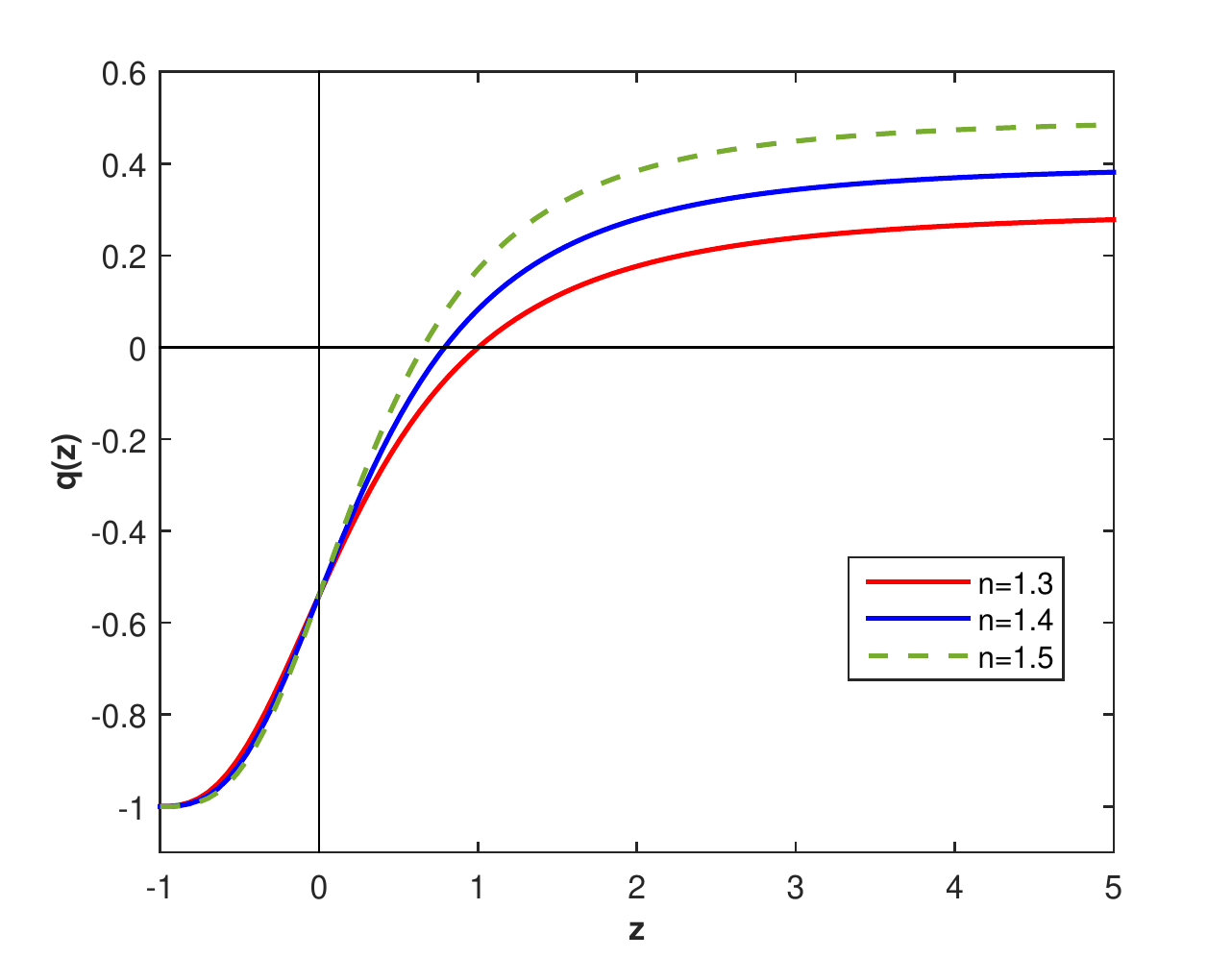}}
\caption{{The plot of the DP $q$ vs. redshift $z$ with $q_{0}=-0.54$.}}
\label{fig2}
\end{figure*}

\begin{equation}
q\left( z\right) =n-1-n\left[ \tanh \left( \sinh ^{-1}\sqrt{\frac{n-\left(
q_{0}+1\right) }{\left( z+1\right) ^{2n}\left( q_{0}+1\right) }}\right) %
\right] ^{2}.  \label{eqn49}
\end{equation}

From Eq. (\ref{eqn45}), it is clear that $q>0$ for $t<\frac{1}{\gamma }\tanh
^{-1}\left( 1-\frac{1}{n}\right) ^{\frac{1}{2}}$ and $q<0$ for $t>\frac{1}{%
\gamma }\tanh ^{-1}\left( 1-\frac{1}{n}\right) ^{\frac{1}{2}}$, and it
predicts the transition phase i.e. $q=0$ at $t=\frac{1}{\gamma }\tanh
^{-1}\left( 1-\frac{1}{n}\right) ^{\frac{1}{2}}$. In \cite{ref42} it is
shown that for $0<n\leq 1$ the model is in the deceleration phase, while for 
$n>1$, the model of Universe exhibits a phase transition from early
decelerating phase to present accelerating phase, which is in good agreement
with the results of recent observations. Thus, we can choose a value of $n$
which gives us the physical behavior of the DP consistent with the
observation data. Fig. \ref{fig2} shows the behavior of the DP in terms of
redshift, in which the parameter $\gamma $ as fixed and three values of the
parameter $n$, especially, $1.4,$ $1.5$ and $1.556$ corresponding to the
values of the transition redshift $z_{tr}=0.57$, $0.63$, and $0.75$,
respectively. The redshift transition values $z_{tr}$ for our cosmological
model are consistent with the observational data \cite{ref43, ref44, ref45}.

Using Eq. (\ref{eqn39}) in (\ref{eqn29}), we get the HDE energy density

\begin{equation}
\allowbreak \rho _{\Lambda }=\frac{3\gamma ^{2}}{n^{2}}\left[ n\beta +\alpha
\coth ^{2}\left( \gamma t\right) -n\beta \coth ^{2}\left( \gamma t\right) %
\right] .  \label{eqn50}
\end{equation}

\begin{figure*}[ht]
\centerline{\includegraphics[scale=0.8]{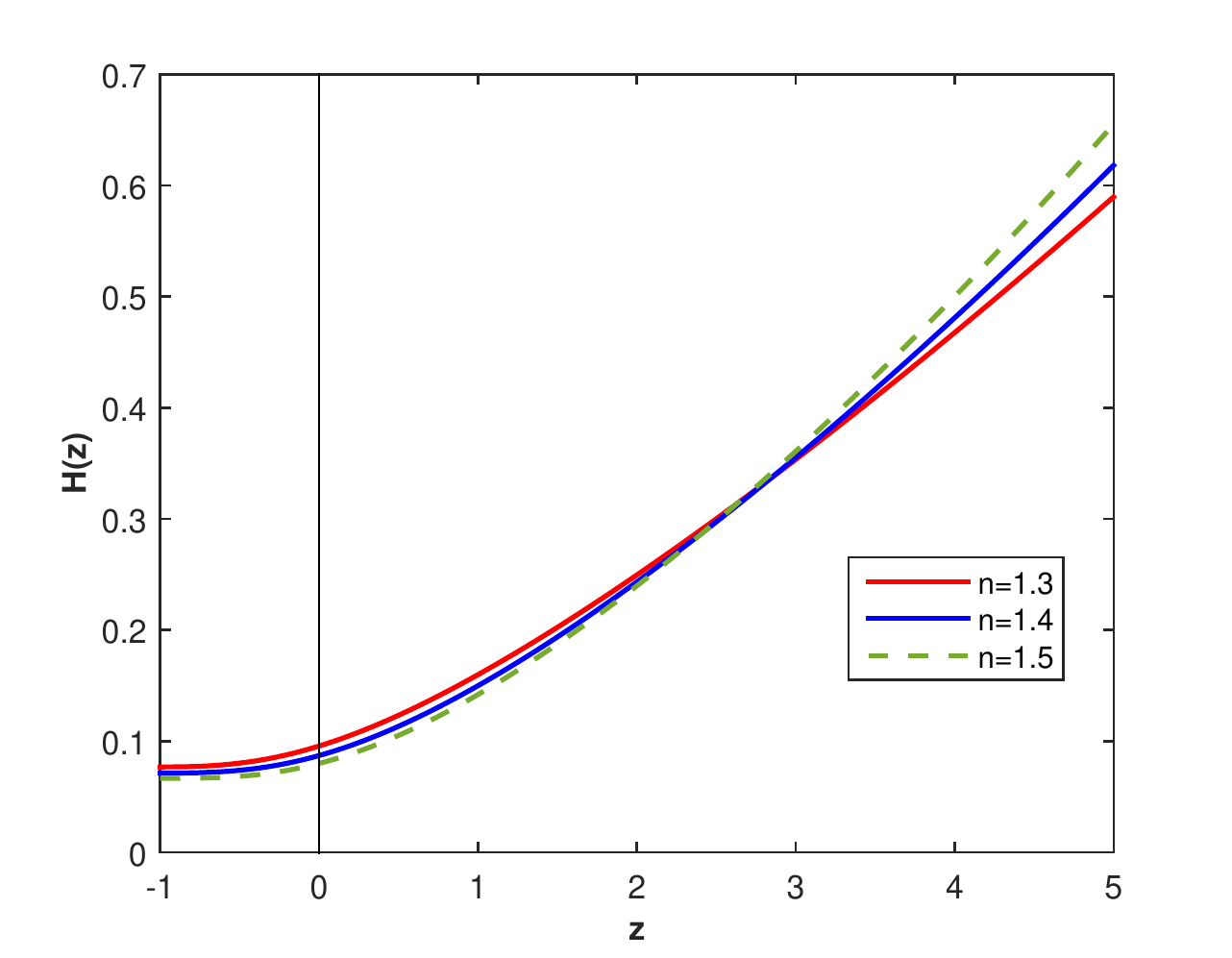}}
\caption{{The plot of the Hubble parameter $H$ vs. redshift $z$ with $%
\protect\gamma =0.1$, $q_{0}=-0.54$.}}
\label{fig3}
\end{figure*}

Again, using Eq. (\ref{eqn39}) in (\ref{eqn27}), we get the matter energy
density

\begin{equation}
\rho _{m}=c\exp \left[ -\frac{3}{n}\left( \ln \left( e^{2\gamma t}-1\right)
-\gamma t\right) \right] ,  \label{eqn51}
\end{equation}%
where $c$ is a constant of integration.

Using Eq. (\ref{eqn39}) in (\ref{eqn30}), we get the EoS parameter of the HDE

\begin{equation}
\omega _{\Lambda }=-1+\frac{2n\left( \coth ^{2}\left( \gamma t\right)
-1\right) \left( \alpha -n\beta \right) }{3n\beta +3\alpha \coth ^{2}\left(
\gamma t\right) -3n\beta \coth ^{2}\left( \gamma t\right) }\allowbreak .
\label{eqn52}
\end{equation}

\begin{figure*}[ht]
\centerline{\includegraphics[scale=1]{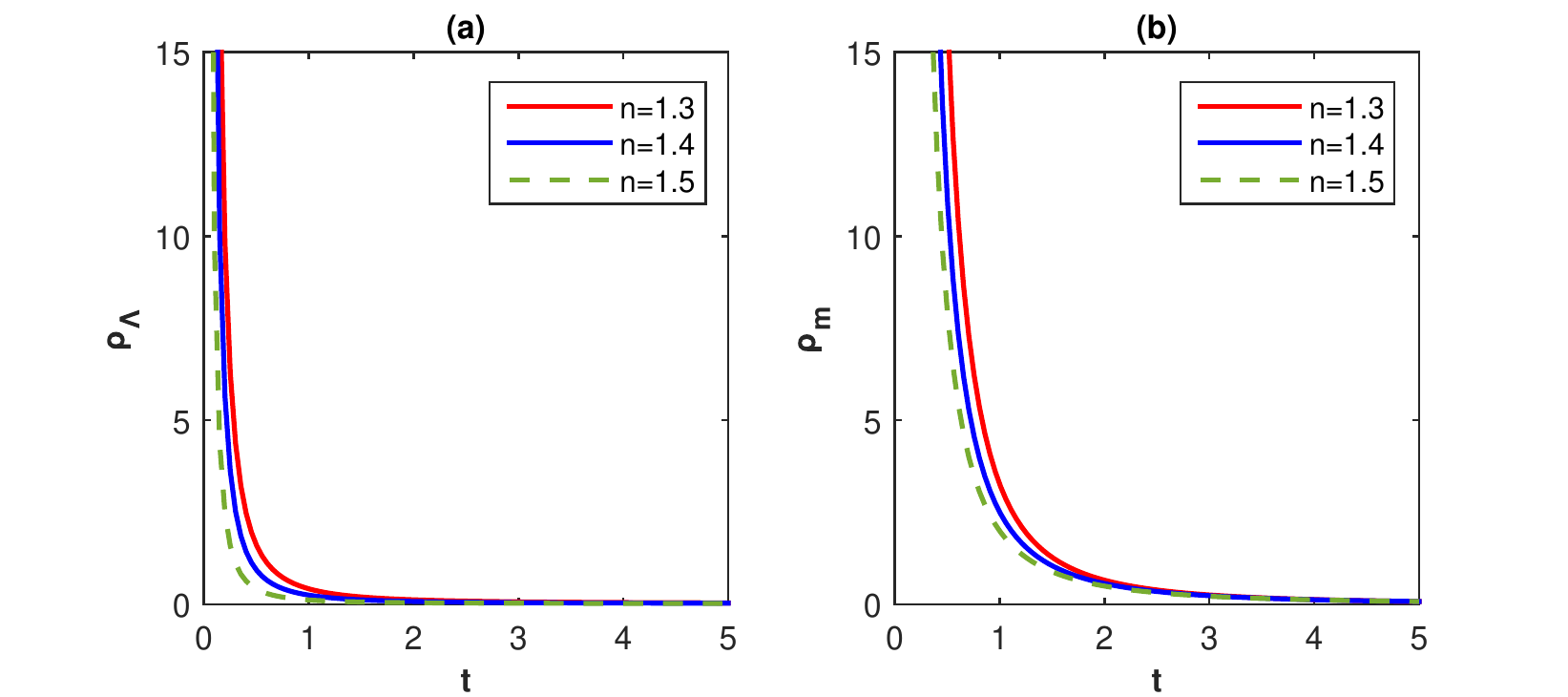}}
\caption{{The plots of (a) HDE density $\protect\rho _{\Lambda }$ vs. time $%
t $ \ and (b) the matter energy density $\protect\rho _{m}$ vs. time $t$
with $\protect\gamma =0.1$, $\protect\alpha =1.2$, $\protect\beta =0.75$ and 
$c=0.08$.}}
\label{fig4}
\end{figure*}

\begin{figure*}[ht]
\centerline{\includegraphics[scale=1]{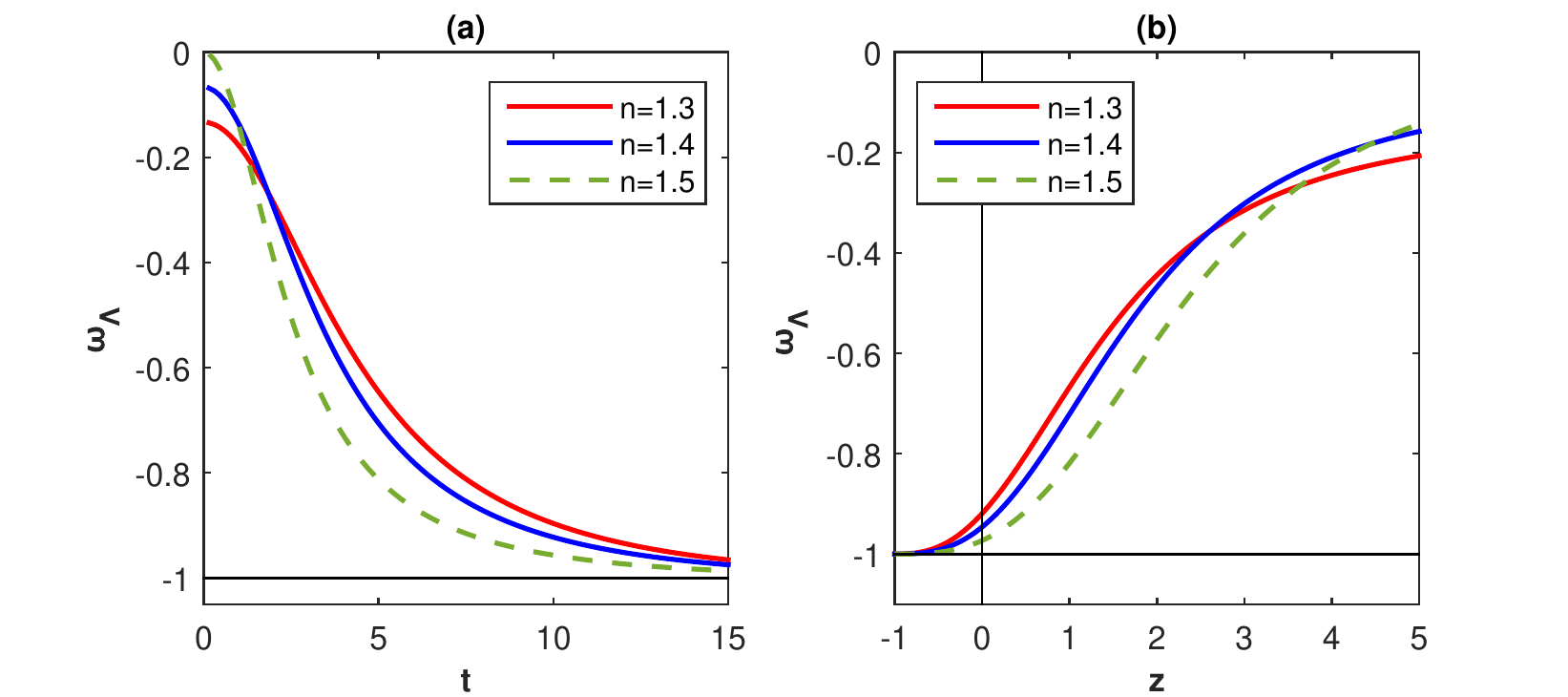}}
\caption{{The plots of (a) the HDE EoS parameter $\protect\omega _{\Lambda }$
vs. time $t$ and (b) HDE EoS parameter $\protect\omega _{\Lambda }$ vs.
redshift $z$ with $\protect\gamma =0.1$, $\protect\alpha =1.2$ and $\protect%
\beta =0.75$.}}
\label{fig5}
\end{figure*}

Fig. \ref{fig4} shows that the energy densities of matter and HDE are
positive decreasing functions of cosmic time. These densities start with an
infinite value at the beginning of cosmic time $t\rightarrow 0$\ and
approach zero at the end time $t\rightarrow \infty $. Fig. \ref{fig5}\ (a)
indicates that the EoS parameter is a decreasing function with cosmic time
for the values of the constant $n>1$, which starts from the quintessence
region $-1<\omega _{\Lambda }<-\frac{1}{3}$, in which remains constant in
this region for the initial time and approaches the value $\omega _{\Lambda
}=-1$ ($\Lambda $CDM model) in the future. From \ref{fig5} (b) the present
values of the EoS parameter corresponding to $n=1.3$, $1.4$ and $1.5$ are $%
\omega _{0}=-0.92$, $-0.94$ and $-0.98$, respectively. These values are in
excellent agreement with the observations \cite{ref46}.

\begin{figure*}[ht]
\centerline{\includegraphics[scale=1]{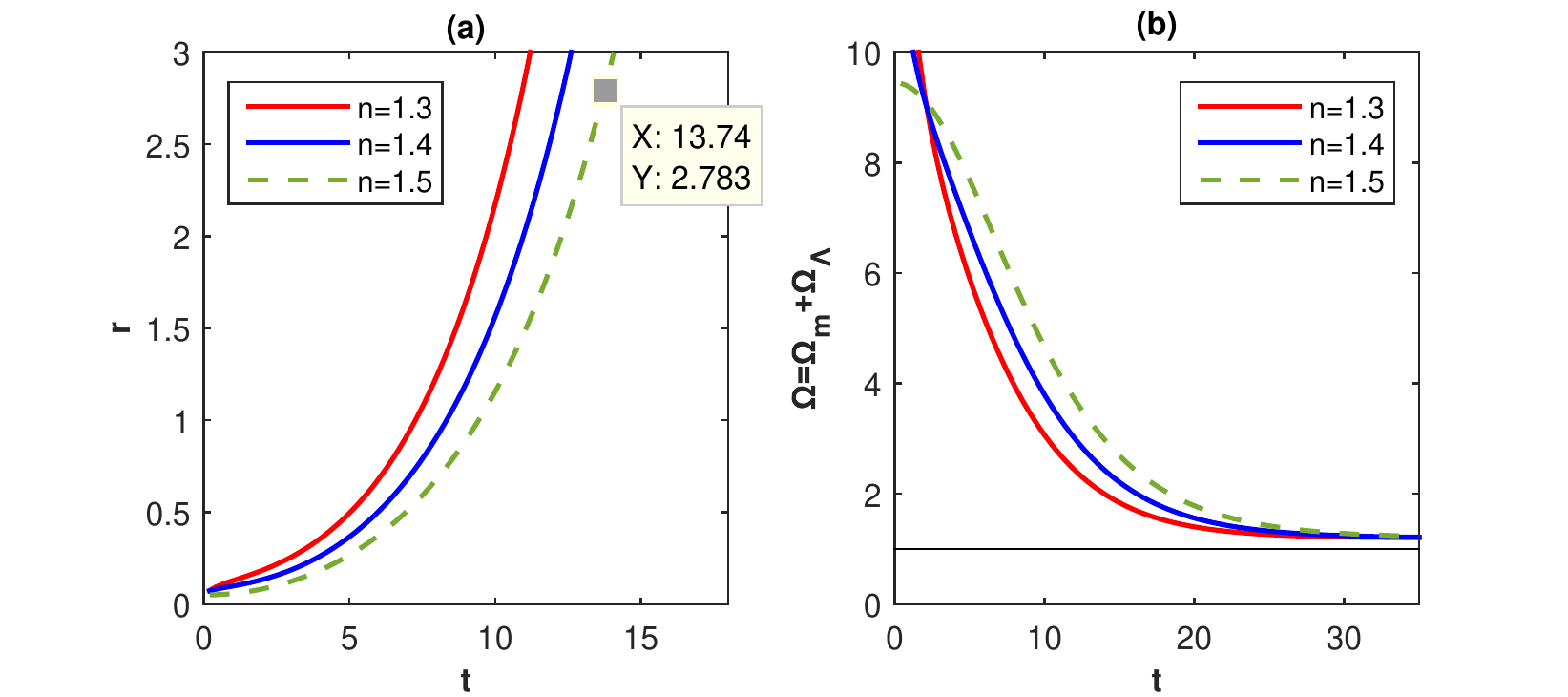}}
\caption{{The plots of (a) the coincidence parameter $r$ vs. time $t$ and
(b) the total energy density parameters $\Omega =\Omega _{m}+\Omega
_{\Lambda }$ vs. time $t$ with $\protect\gamma =0.1$, $\protect\alpha =1.2$, 
$\protect\beta =0.75$ and $c=0.08$.}}
\label{fig6}
\end{figure*}

Let $r$ be the coincidence parameter and defined as $r=\frac{\rho _{\Lambda }%
}{\rho _{m}}$. Hence, by using Eqs. (\ref{eqn50}) and (\ref{eqn51}) the
coincidence parameter becomes

\begin{equation}
r=\frac{\rho _{\Lambda }}{\rho _{m}}=\frac{\frac{3\gamma ^{2}}{n^{2}}\left[
n\beta +\alpha \coth ^{2}\left( \gamma t\right) -n\beta \coth ^{2}\left(
\gamma t\right) \right] }{c\exp \left[ -\frac{3}{n}\left( \ln \left(
e^{2\gamma t}-1\right) -\gamma t\right) \right] }.  \label{eqn53}
\end{equation}

Fig. \ref{fig6}\ (a) indicates the behavior of the coincidence parameter $r$
as a function of cosmic time $t$. From the figure, the current value of the
coincidence parameter, i.e. $t_{0}=13.798$ \textit{GYs} is consistent with
the current value extracted from the observation data \cite{ref46}. Further,
it is useful to use yet another notation, the abundances, also called the
density parameters, it represents the proportion of each component in the
Universe. The total energy density parameter i.e. $\Omega =\Omega
_{m}+\Omega _{\Lambda }$ takes three values: $\Omega >1$, $\Omega =1$, $%
\Omega <1$ correspond to the open, flat, and closed Universe, respectively.
The matter density parameter $\Omega _{m}$\ and HDE density parameter $%
\Omega _{\Lambda }$\ are defined by

\begin{equation}
\Omega _{m}=\frac{\rho _{m}}{3H^{2}}\text{ \ \ and \ \ }\Omega _{\Lambda }=%
\frac{\rho _{\Lambda }}{3H^{2}}.  \label{eqn54}
\end{equation}

Using (\ref{eqn39}), (\ref{eqn50}), (\ref{eqn51}) and (\ref{eqn54}) we get
the total density parameter as

\begin{equation}
\Omega =\Omega _{m}+\Omega _{\Lambda }=\frac{cn^{2}}{3\gamma ^{2}}\frac{\exp %
\left[ -\frac{3}{n}\left( \ln \left( e^{2\gamma t}-1\right) -\gamma t\right) %
\right] }{\coth ^{2}\left( \gamma t\right) }+n\beta \left( \frac{1}{\coth
^{2}\left( \gamma t\right) }-1\right) +\alpha .  \label{eqn55}
\end{equation}

Fig. \ref{fig6} (b) represents the evolution of the total energy density
parameter as a function of cosmic time $t$, and it appears that its value is
large in the first era of the Universe, while begin to approach $\Omega \sim
1$ in the last era of Universe, which causes our cosmological model to
predict a flat Universe at a later time, as recent astronomical observations
indicate.

\section{Jerk parameter}

As it is known in the literature, the jerk parameter is one of the
fundamental physical quantities to describe the dynamics of the Universe.
The Jerk parameter is a dimensionless third derivative of the scale factor $a
$ with respect to cosmic time $t$ and is defined as \cite{ref47, ref48}

\begin{equation}
j=\frac{\overset{...}{a}}{aH^{3}}.  \label{eqn56}
\end{equation}

Eq. (\ref{eqn56}) can be written in terms of the deceleration parameter $q$
as

\begin{equation}
j=q+2q^{2}-\frac{\overset{.}{q}}{H}.  \label{eqn57}
\end{equation}

Using Eqs. (\ref{eqn39}) and (\ref{eqn45}), the jerk parameter for our
cosmological model is

\begin{equation}
j=1+n\left( 2n-3\right) \sec h\left( \gamma t\right) ^{2}.  \label{eqn58}
\end{equation}

To study the behavior of the jerk parameter $j$, it is better to express in
terms of redshift $z$

\begin{equation}
j\left( z\right) =1+\frac{n\left( 2n-3\right) }{1+2.17391\left(
n-0.46\right) \left( 1+z\right) ^{-2n}}.  \label{eqn59}
\end{equation}

\begin{figure*}[ht]
\centerline{\includegraphics[scale=0.8]{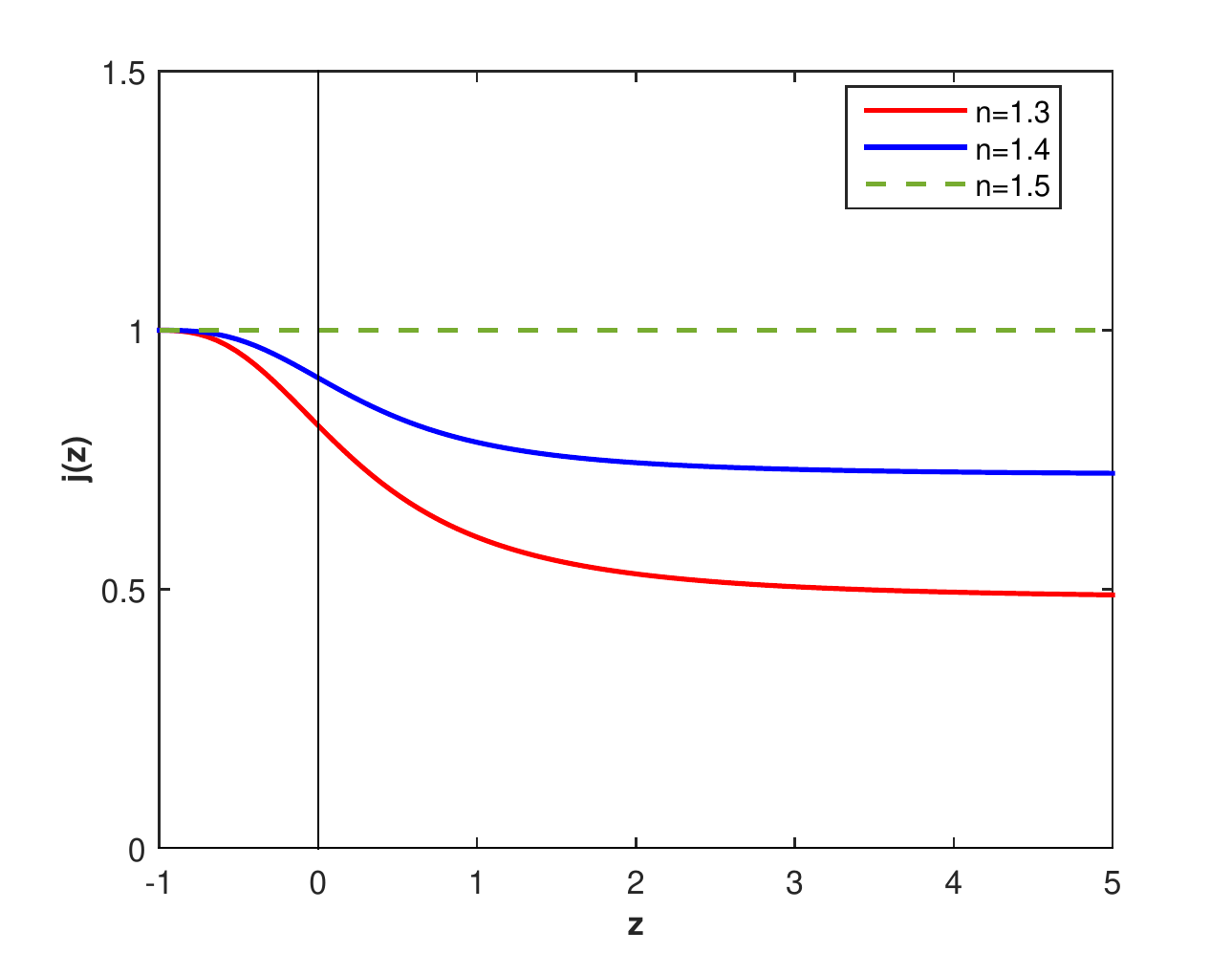}}
\caption{{The plot of the jerk $j $ parameter vs. $z$.}}
\label{fig7}
\end{figure*}

For the $\Lambda $CDM model, the value of the jerk parameter is $j=1$.
According to the $\Lambda CDM$ model, the Universe shifts from the early
deceleration phase to the current acceleration phase with a positive jerk
parameter $j_{0}>0$ and a negative DP $q_{0}<0$. From Fig. \ref{fig7} it is
clear that the jerk parameter remains positive in various cases for $n>1$
and approaches $1$ later. The current jerk parameter value $j_{0}$ is
positive. Thus, for $n=1.3$ and $1.4$, our cosmological model can be
expected to adopt the behavior of another DE model instead of the $\Lambda $%
CDM model, while for $n=1.5$ our cosmological model is similar to the $%
\Lambda $CDM model.

\section{Conclusions}

In this work, we investigated a spatially homogeneous and anisotropic
Bianchi Type-I Universe with two fluids as the content of the Universe:
matter and holographic dark energy (HDE) in the framework of general
relativity. We considered a scale factor as a hyperbolic function,
specifically, $a\left( t\right) =\sinh ^{\frac{1}{n}}\left( \gamma t\right) $%
, where $\gamma $ and $n>0$ are arbitrary constants, which gives us a
time-dependent deceleration parameter. Then we derived the Einstein's field
equations for Bianchi Type-I Universe. We found the exact solutions for our
cosmological model. Further, to obtain a Universe moving from early
decelerating phase to present accelerating phase, we choose the value of $n>1
$ \cite{ref42}. In addition, we have investigated the behavior of
anisotropic parameter and deceleration parameter for the for the three
values of model parameters of $n$ i.e. $n=1.3$, $1.4$ and $1.5$. The
evolution of the deceleration parameter in Fig. \ref{fig1} indicates that
our cosmological model contains a transition from the early anisotropic
Universe to the current isotropic Universe as DE starts to dominate the
energy density of the Universe and the deceleration parameter in Fig. \ref%
{fig2} show a phase transition from early decelerating phase to current
accelerating phase, which is in good concurrence with the results of recent
observations. The values of the transition redshift corresponding to the
three values of model parameter $n$ are $z_{tr}=0.57$, $0.63$, and $0.75$,
respectively.

In addition, we have investigated the behavior of the energy densities of
matter and HDE for the for the three values of model parameter $n$. From
Fig. \ref{fig4} we observed that the energy densities of matter and HDE are
positive decreasing functions of cosmic time. They start with an infinite
value at the beginning of cosmic time $t\rightarrow 0$\ and approach zero at
the end time $t\rightarrow \infty $. These results are consistent with the
expansion of the Universe. Additional results of our cosmological model show
that under certain conditions, the ratio of the HDE density to the energy
density of matter $r$ increases with the expansion of the Universe, i.e.
that the Universe moves from the era of the domination of matter to the era
of the domination of DE. This is good for the problem of cosmic coincidence.
Further, the equation of state (EoS) parameter presented in Fig. \ref{fig5}
indicates that the two fluids of the model behaves like quintessence dark
energy in present. The value of the EoS parameter at present epoch $z=0$ for
dark energy obtained by Planck collaboration is $\omega _{0}=-1.026\pm 0.032$
\cite{ref46}. In our analysis, we found $\omega _{0}=-0.92$, $-0.94$ and $%
-0.98$ corresponding to the three values of model parameter $n$,
respectively, which is in good agreement with Planck measurements. Finally,
the evolution of the total density parameter presented in Fig. \ref{fig6}
indicates that our cosmological is close to $1$ in the current time, which
leads to a flat Universe. We also found that the jerk parameter is similar
to $\Lambda $CDM model in the future.

\section*{Acknowledgments}

We are very much grateful to the honorary referee and the editor for the
illuminating suggestions that have significantly improved our work in terms
of research quality and presentation.\newline

\textbf{Data availability} There are no new data associated with this article%
\newline

\textbf{Declaration of competing interest} The authors declare that they
have no known competing financial interests or personal relationships that
could have appeared to influence the work reported in this paper.\newline


\begin{thebibliography}{99}
\bibitem{ref1} Perlmutter S., Aldering G., della Valle M., et al. 1998, 
\textit{Nature}, \textbf{391}, 51-54.

\bibitem{ref2} Perlmutter S., Aldering G., Goldhaber G., et al. 1999, 
\textit{The Astrophysical Journal}, \textbf{517}, 565-586.

\bibitem{ref3} Filippenko A. v., Riess A. G. 1998, \textit{Physics Reports}, 
\textbf{307}, 31-44.

\bibitem{ref4} Tonry J. L., Schmidt B. P., Barris B., et al. 2003, \textit{%
The Astrophysical Journal}, \textbf{594}, 1-24.

\bibitem{ref5} Clocchiatti A., Schmidt B. P., Filippenko A. v., et al. 2006, 
\textit{The Astrophysical Journal,} \textbf{642}, 1-21.

\bibitem{ref6} de Bernardis P., Ade P. A. R., Bock J. J., et al. 2000, 
\textit{Nature}, \textbf{404}, 955-959.

\bibitem{ref7} Hanany S., Ade P., Balbi A,. et al. 2000, \textit{The
Astrophysical Journal}, \textbf{545}, L5-L9.

\bibitem{ref8} Blake C., Kazin E. A., Beutler F., et al. 2011, \textit{%
Monthly Notices of the Royal Astronomical Society}, \textbf{418}, 1707-1724.%

\bibitem{ref9} Padmanabhan N., Xu X., Eisenstein D. J., et al. 2012, \textit{%
Monthly Notices of the Royal Astronomical Society}, \textbf{427}, 2132-2145.%

\bibitem{ref10} Anderson L., Aubourg E., Bailey S., et al. 2012, ,\textit{\
Monthly Notices of the Royal Astronomical Society}, \textbf{427}, 3435-3467.%

\bibitem{ref11} Hinshaw G., Larson D., Komatsu E., et al. 2013,\textit{\
Astrophysical Journal, Supplement Series}, \textbf{208}, 19.

\bibitem{ref12} Zlatev I., Wang L., Steinhardt P. J. 1999, \textit{Physical
Review Letters} \textbf{82}, 896-899.

\bibitem{ref13} Carroll S. M. 1998, \textit{Physical Review Letters} \textbf{%
81}, 3067.

\bibitem{ref14} Turner M. S. 2002, \textit{A Spacetime Odyssey}, 180-196.%

\bibitem{ref15} Caldwell R. R. 2002, \textit{Physics Letters, Section B:
Nuclear, Elementary Particle and High-Energy Physics}, \textbf{545}, 23-29.%

\bibitem{ref16} Chiba T., Okabe T., Yamaguchi M. 2000, \textit{Physical
Review D}, \textbf{62}, 8.

\bibitem{ref17} Padmanabhan T. 2002, \textit{Physical Review D} \textbf{66},
021301.

\bibitem{ref18} Kamenshchik A., Moschella U., Pasquier V. 2001, P\textit{%
hysics Letters, Section B: Nuclear, Elementary Particle and High-Energy
Physics}, \textbf{511}, 265-268.

\bibitem{ref19} Myrzakulov R. 2011,\textit{\ The European Physical Journal C}%
, \textbf{71}, 1-8.

\bibitem{ref20} Linder E. v. 2010, \textit{Physical Review D}, \textbf{81},
127301.

\bibitem{ref21} De Laurentis M., Paolella M., Capozziello S.2015, \textit{%
Physical Review D}, 91.

\bibitem{ref22} Harko T., Lobo F. S. N., Nojiri S., Odintsov S. D. 2011, 
\textit{Physical Review D}, \textbf{84}, 1-11.

\bibitem{ref23} Hooft G 't. 1993, arXive gr-qc/9310026.

\bibitem{ref24} Fischler W., Susskind L. 1998, \textit{Holography and
Cosmology}.

\bibitem{ref25} Wang S., Wang Y., Li M. 2017, \textit{Physics Reports}, 
\textbf{696}, 1-57.

\bibitem{ref26} Li M. 2004, \textit{Physics Letters, Section B: Nuclear,
Elementary Particle and High-Energy Physics}, \textbf{603}, 1-5.

\bibitem{ref27} Granda L. N., Oliveros A. 2008, \textit{Physics Letters B}, 
\textbf{669}, 275-277.

\bibitem{ref28} Sarkar S. 2014, \textit{Astrophysics and Space Science}, 
\textbf{349}, 985-993.

\bibitem{ref29} Sarkar S. 2016, \textit{International Journal of Theoretical
Physics}, \textbf{55}, 481-494.

\bibitem{ref30} Sarkar S., Mahanta C. R. 2013, \textit{International Journal
of Theoretical Physics}, \textbf{52}, 1482-1489.

\bibitem{ref31} Samanta G. C. 2013, \textit{International Journal of
Theoretical Physics}, \textbf{52,} 4389-4402.

\bibitem{ref32} Dubey V. C., Srivastava S., Sharma U. K., Pradhan A. 2019, 
\textit{Pramana - Journal of Physics}, \textbf{93}, 1-10.

\bibitem{ref33} Koussour M., Bennai M. 2021, \textit{International Journal
of Geometric Methods in Modern Physics}, \textbf{19}, 03.

\bibitem{ref34} Koussour M., Bennai M. 2022, \textit{International Journal
of Modern Physics A}, 37.

\bibitem{ref35} Koussour M., Bennai M. 2022, \textit{Afrka Matematika}, 
\textbf{33}, 1-16.

\bibitem{Xu} Xu, L. 2009, \textit{Journal of Cosmology and Astroparticle
Physics}, \textbf{09}, 016.

\bibitem{Chen} Chen, S., Jing, J. 2009, \textit{Physics Letters B}, \textbf{%
679}, 2.

\bibitem{ref36} Chawla C., Mishra R. K., Pradhan A. 2012, \textit{The
European Physical Journal Plus} \textbf{127}, 1-16.

\bibitem{ref37} Pradhan A. 2014, \textit{Indian Journal of Physics}, \textbf{%
88}, 215-223.

\bibitem{Esmaeili} Esmaeili, F. M., Mishra, B. 2018. \textit{Journal of
Astrophysics and Astronomy}, \textbf{39}, 5.

\bibitem{ref38} Pradhan A., Tiwari R. K., Beesham A., Zia R. 2019, \textit{%
The European Physical Journal Plus}, \textbf{134}, 1-18.

\bibitem{Singh} Singh, G. P., Lalke, A. R. 2022, \textit{Indian Journal of
Physics}, 1-12.

\bibitem{ref39} Ahmed N., Pradhan A. 2014,\textit{\ International Journal of
Theoretical Physics,} \textbf{53}, 289-306.

\bibitem{ref40} MacCallum M. A. H. 1971, \textit{Communications in
Mathematical Physics}, \textbf{20}, 57-84.

\bibitem{ref41} Mamon A. al., Das S. 2017, \textit{The European Physical
Journal C}, \textbf{77}, 1-9.

\bibitem{ref42} Nagpal R., Singh J. K., Beesham A., Shabani H. 2019, \textit{%
Annals of Physics} \textbf{405}, 234-255.

\bibitem{ref43} Capozziello S., Farooq O., Luongo O., Ratra B. 2014, \textit{%
Physical Review D}, \textbf{90}, 044016.

\bibitem{ref44} Capozziello S., Luongo O., Saridakis E. N. 2015, \textit{%
Physical Review D}, \textbf{91}, 124037.

\bibitem{ref45} Farooq O., Madiyar F. R., Crandall S., Ratra B. 2017, 
\textit{The Astrophysical Journal}, \textbf{835}, 26.

\bibitem{ref46} Aghanim N., Akrami Y., Ashdown M., et al. 2020, \textit{%
Astronomy \& Astrophysics}, \textbf{641}, A6.

\bibitem{ref47} Visser M. 2005, \textit{General Relativity and Gravitation}, 
\textbf{37}, 1541-1548.

\bibitem{ref48} Rapetti D., Allen S. W., Amin M. A., Blandford R. D. 2007, 
\textit{Monthly Notices of the Royal Astronomical Society}, \textbf{375},
1510-1520.
\end{thebibliography}
\end{document}